\documentclass[letter]{aa} 

\usepackage{graphicx}
\usepackage{txfonts}
\usepackage{hyperref}
\begin{document}

   \title{Simulating the pericentre passage of the Galactic centre star S2}

   \author{M.~Schartmann\inst{1,2}\thanks{E-mail: schartmann@mpe.mpg.de}
          \and
          A.~Burkert\inst{1,2}
          \and
          A.~Ballone\inst{3}
          }
   \institute{Universit\"ats-Sternwarte M\"unchen, Scheinerstra\ss e 1, D-81679 M\"unchen, Germany\\\
         \and
             Max-Planck-Institut f\"ur extraterrestrische Physik, Postfach 1312, Giessenbachstr. 1, D-85741 Garching, Germany\\
         \and
             INAF-Osservatorio Astronomico di Padova, Vicolo dell'Osservatorio 5, I–35122, Padova, Italy\\
             }
   \date{Received 4 April 2018; Accepted 30 July 2018}


  \abstract
   {Our knowledge of the density distribution of the accretion flow around Sgr~A* -- the
    massive black hole (BH) at our Galactic centre (GC) -- relies on two measurements only: one at a
    distance of a few Schwarzschild radii ($R_\mathrm{s}$) and one at roughly $10^5~R_\mathrm{s}$, which are
    usually bridged by a power law, which is backed by magnetohydrodynamical simulations.
    The so-called S2 star reached its closest approach to the massive BH
     at around 1500~$R_\mathrm{s}$ in May 2018.
     It has been proposed that the interaction of its stellar wind with the high-density accretion flow at this distance from Sgr~A*
     will lead to a detectable, month-long X-ray flare.}
   {Our goal is to verify whether or not the S2 star wind can be used as a diagnostic tool to infer the properties of the
    accretion flow towards Sgr~A* at its pericentre (an unprobed distance regime), putting important constraints on BH accretion flow models.}
   {We run a series of three-dimensional adaptive mesh refinement simulations with the help of the
    {\sc Ramses} code which include the realistic treatment of the interaction of S2's stellar wind
    with the accretion flow along its orbit and -- apart from hydrodynamical and thermodynamical effects -- include
    the tidal interaction with the massive BH. These are post-processed to derive the
    X-ray emission in the observable 2-10\,keV window.}
   {No significant excess of X-ray emission from Sgr~A* is found for typical
    accretion flow models.
    A measurable excess is produced for a significantly increased density of the accretion flow.
    This can, however, be ruled out for standard power-law accretion flow models as in this case the
    thermal X-ray emission without the S2 wind interaction would already exceed the observed
    quiescent luminosity.
    Only a significant change of the wind parameters
    (increased mass loss rate and decreased wind velocity)
    might lead to an (marginally) observable X-ray flaring event.}
   {Even the detection of an (month-long) X-ray flare during the pericentre passage of the S2 star
     would not allow for strict constraints to be put on the accretion flow around Sgr~A* due to the
     degeneracy caused by the dependence on multiple parameters (of the accretion flow
     model as well as the stellar wind).
   }
   \keywords{Galaxy: centre --
             Stars: winds, outflows --
             Hydrodynamics --
             Accretion, accretion disks --
             Black hole physics --
             Gravitation
               }

   \maketitle
%

\section{Introduction}
\label{sec:introduction}

The massive central black hole (BH) makes the Galactic centre (GC) a unique laboratory to
study gas and stellar dynamics in an extreme environment.
With a visual extinction of $A_\mathrm{V}\approx 30$\,mag
\citep{Becklin_68,Rieke_89},
corresponding to a column density of $1.6\times 10^{23}$\,cm$^{-2}$\citep{Ponti_17},
the obscuring medium becomes partially transparent to X-rays with energies above 2\,keV.
In the spectral window between 2 and 10\,keV, \citet{Baganoff_03} found a source with
a full width at half maximum
(FWHM) of 1.4\,as ($\approx 0.06\,$pc) and an absorption corrected luminosity of $2.4 \times 10^{33}$ erg\,s$^{-1}$
(with roughly a factor of two uncertainty)
at the position of Sgr~A*, the compact nonthermal radio source \citep{Balick_74,Backer_96}
associated with the massive BH at the dynamical centre of the Galaxy.
Assuming that this emission mainly arises from thermal X-rays emitted by a hot accretion
flow towards Sgr~A*, \citet{Baganoff_03} fit an absorbed optically thin thermal plasma model,
finding that the average electron density inside this region is
$n_e \approx 130\,\mathrm{cm}^{-3}$ with a temperature of $k_\mathrm{B}T_\mathrm{e} \approx 2$~keV.
The fuel for the accretion flow at the Bondi radius is thought to be provided by winds of a
cluster of young, luminous stars
\citep{Krabbe_95,Najarro_97,Coker_97,Quataert_99,Paumard_01}.
Radio observations, by measuring the polarization (from Faraday rotation)
and the size of the emission, allow for  the density
to be roughly constrained at a distance of a few $R_\mathrm{s}$
\citep{Quataert_00b,Bower_03,Marrone_07,Doeleman_08}.
In the region in between, the density and temperature distributions of
the accretion flow are simply estimated with the help of power-law models (Sect.~\ref{sec:accflow}),
backed by magnetohydrodynamical simulations.
An attempt to constrain the density distribution close to the pericentre position of the
G2 cloud (about 1500\,$R_\mathrm{s}$) was made by \citet{Plewa_17}. By means of fitting test-particle simulations to the
observations, they find an electron number density of around $10^3\,\mathrm{cm}^{-3}$.
Within the same distance regime,
the S-star cluster is located
\citep{Schoedel_03,Ghez_05a,Eisenhauer_05,Gillessen_09a}. Its best observed member
is the S2 star \citep[e.g.~][]{Gillessen_09b}, which is on a $\sim$ 16 year orbit and
reached its pericentre passage (also at around 1500\,$R_\mathrm{s}$
distance) in May 2018. It is an early B-dwarf of spectral type B0-B2.5 V with a
stellar wind with an estimated velocity $v_{\mathrm{w}}\sim1000$\,km s$^{-1}$ and a
mass-loss rate of $\dot{M}_{\mathrm{w}}\lesssim 3\times 10^{-7}$\,M$_{\sun}$yr$^{-1}$ \citep{Martins_08}.
It has been suggested to use the X-ray emission from the interaction of S2's
stellar wind with the accretion flow during pericentre passage to put
constraints on its density
structure in this so-far unprobed distance regime \citep{Giannios_13,Christie_16}.
With the help of analytical estimates of the bow shock interaction with
an already present gas distribution around Sgr~A*, \citet{Giannios_13}
estimated the X-ray emission to be significant around pericentre passage
and to be sensitive to the exponent of the assumed radial (power law) profile of the
accretion flow.
An additional contribution to the X-ray emission during S2's pericentre
passage (which we neglect in this
letter) could arise from Compton-upscattered optical/UV photons emitted from the star as was
suggested by \citet{Nayakshin_05}.
The radio emission has been analytically estimated by \citet{Crumley_13},
\citet{Zajacek_16} and \citet{Ginsburg_16}.
Two-dimensional (2D) numerical hydro-simulations were presented by \citet{Christie_16}
to estimate the increase in thermal X-ray emission during the
pericentre passage of the S2 star.
For their choice of parameters they find that S2's pericentre passage leads to a
detectable roughly one month long X-ray flare emission (Sect.~\ref{sec:comp_theo}).
However, given the complex interplay of the hydrodynamical, thermodynamical, and
gravitational interaction
of the stellar wind shells with the ambient medium, neither the analytical estimates nor the
2D simulations are expected to capture the full complexity of the problem.
Three-dimensional (3D) simulations including the evolution along the orbit are necessary and
are presented in this letter.
We describe the simulation setup and analysis in
Sect.\,\ref{sec:simsetup}, discuss the results in Sect.\,\ref{sec:results_discussion},
and conclude in Sect.~\ref{sec:conclusions}.

\section{Simulation setup and analysis}
\label{sec:simsetup}

We adopt the orbital elements of the S2 star
from Table~3 of \citet{Gillessen_17} with the
solar distance to Sgr~A* of 8.13\,kpc and
a mass of the central BH of $4.1 \times 10^6$\,M$_{\sun}$.
We start the mechanically implemented wind ($v_{\mathrm{w}}=1000$\,km s$^{-1}$
and $\dot{M}_{\mathrm{w}} = 3\times 10^{-7}$\,M$_{\sun}$yr$^{-1}$ for the
standard simulation) in apo centre and
follow its hydrodynamical, thermodynamical (adiabatic with solar metallicity
cooling), and gravitational interaction using the adaptive mesh
refinement code {\sc Ramses} \citep{Teyssier_02}.

\subsection{The accretion flow around Sgr~A*}
\label{sec:accflow}

The quiescent galactic nucleus of the Milky Way is found to be underluminous for
its estimated accretion rate with respect to~standard thin accretion disc models \citep{Baganoff_03}.
In order to account for this misbalance, radiatively inefficient, optically thin, geometrically thick
accretion flow models have been proposed, featuring high gas temperatures and low densities that
are often described by power-law radial profiles and that give rise to thermal
Bremsstrahlung emission.
Various exponents for the radial power-law density distribution have been suggested
within the range of $\beta=-1.5$ and $-0.5$:
the Bondi solution is described by roughly $-1.5$ and so-called convection-dominated accretion flow (CDAF) models \citep{Quataert_00a}
predict $-0.5$. A value of -1.0 has been found in simulations \citep[e.g.~][]{McKinney_12,Narayan_12}
and by analytical, so-called radiatively inefficient accretion flow (RIAF) solutions \citep{Yuan_03}. The latter has been widely used to describe
the interaction of the G2 cloud with the accretion flow
\citep{Gillessen_12,Burkert_12,Schartmann_15,Ballone_13}.
We therefore chose this solution for our standard model,
employing the same resetting technique as in \citet{Schartmann_15}, due to its
instability to convection.
Additionally, we test other exponents of the power law density distribution, as
the most recent Chandra observations favor a less steep profile with an exponent close to -0.5 \citep{Wang_13,Roberts_17}.
We normalise the accretion flow models so as to obtain the observed quiescent 2-10\,keV thermal emission
within a radius of $0.7\,\arcsec$ -- the intrinsic size of the Sgr\,A* emission
found by \citet{Baganoff_03}.
For our standard model, this results in a density distribution of
\begin{eqnarray}
\label{equ:densdistribution}
\rho = 2.2\times 10^{-20}\, \left(\frac{r}{1.8\times 10^{15}\,\mathrm{cm}} \right)^{-1}\,\mathrm{g} \, \mathrm{cm}^{-3},
\end{eqnarray}
where $r$ is the distance to Sgr~A*.
Despite neglecting magnetic fields and rotation and setting it up in
hydrostatic equilibrium, we refer to it as the
``accretion flow'' in the following. The simulations range from an inner radius of
$r_{\mathrm{in}} = 1.4 \times 10^{-4}\,$pc to the edge of a cubic box with width of
roughly $3.3 \times 10^{-2}\,$pc, reaching a minimum cell size of $4\,\muup$pc.

\subsection{Modelling the X-ray emission}
\label{sec:xray_modelling}

The intrinsic thermal X-ray emission properties of the hot gas are calculated
using the Astrophysical Plasma Emission Code (APEC) model \citep{Smith_01} within the AtomDB \citep{Foster_12} version 3.0.9.
An optically thin, thermal plasma in collisional ionisation equilibrium (CIE) is
assumed. The resulting X-ray emission per cell is then projected onto the orbital
plane and images and light curves in the 2-10\,keV window are calculated.
We subtract the X-ray emission of our initial condition from the one obtained
in the individual time steps in order to single out the excess
emission due to the interaction of the stellar wind with the accretion flow.

\section{Results and discussion}
\label{sec:results_discussion}

\subsection{The wind / accretion flow interaction}
\label{sec:dens_evol}

\begin{figure*}
\centering
\includegraphics[width=\hsize]{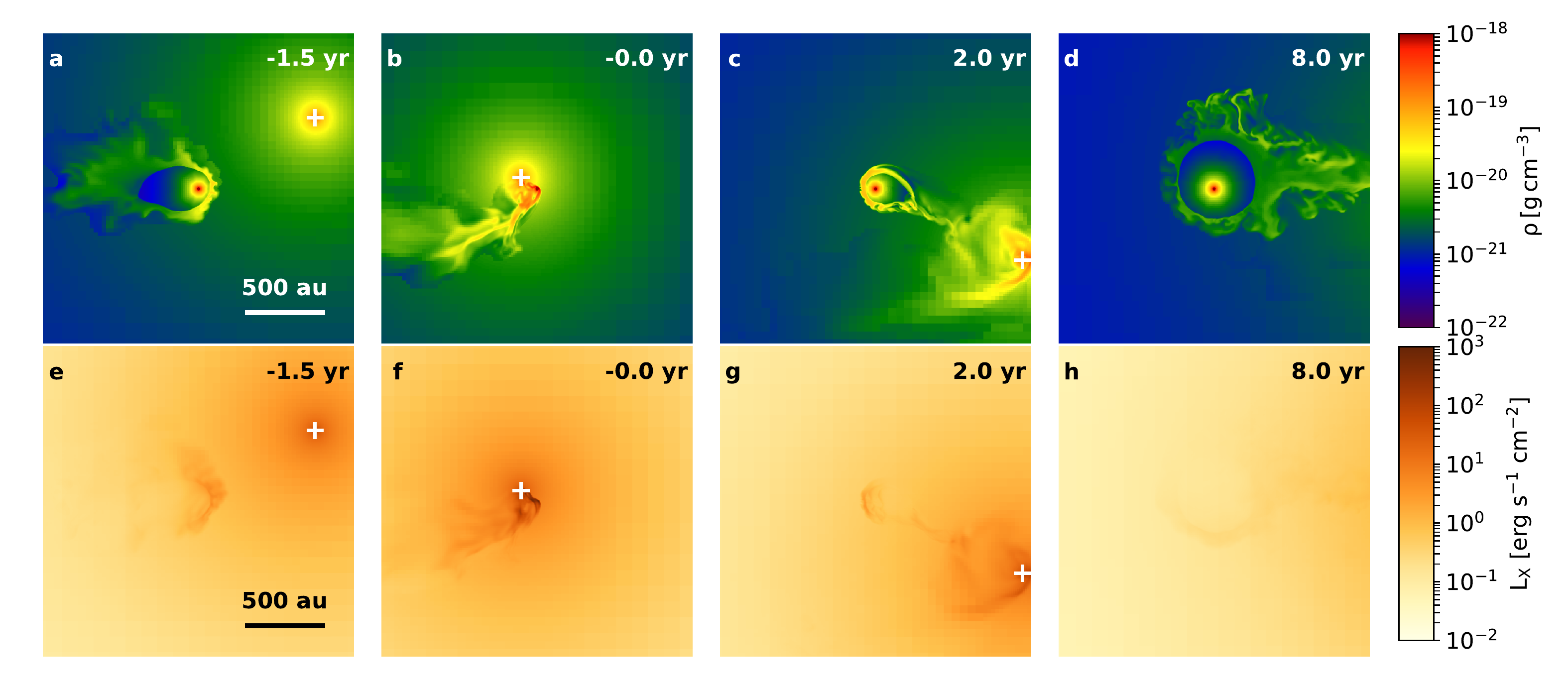}
\caption{Time evolution (given relative to the epoch of pericentre passage) of the standard model. Shown are cuts through
         the density distribution (centred on S2) within the orbital plane (a-d) and
         X-ray 2-10\,keV emission maps (perpendicular to the orbital plane), including the emission of the accretion flow (e-h).
         The white cross depicts the location of the BH and is located outside (to the lower right) of the shown region in panels d and h.
         }
\label{fig:letter_s2star_evolution}
\end{figure*}

Snapshots of the density evolution of our standard model are shown in Fig.~\ref{fig:letter_s2star_evolution}a-d.
We start the simulation itself and the stellar wind at the apocentre.
As the stellar velocity
is relatively small ($v_\mathrm{*}^{\mathrm{apo}}\approx 469\,\mathrm{km}\,\mathrm{s}^{-1}$) at this position,
the stellar wind has time to expand and reach its stagnation radius, where the ram pressure of the
wind is balanced by the thermal pressure of the ambient medium. This expansion phase
excites Rayleigh-Taylor instabilities that break up the contact discontinuity and
start to transform the shocked wind shell into a structure with filaments and
mushroom-shaped fingers pointing
outwards, but leaving the inner shock as a sharp transition towards the free streaming
wind (Fig.~\ref{fig:letter_s2star_evolution}a).
Due to the very high pressure in the GC, the outer shock is very weak.
Moving into the higher-density (and pressure) inner regions of the
BH accretion flow, the free wind region decreases again and its spherical shape is turned
into a drop-like shape, as the ram pressure due to the motion of the star becomes
important and adds to the ambient thermal pressure in the aforementioned equilibrium.
The latter causes the asymmetric location of the star in the free-wind region (Fig.~\ref{fig:letter_s2star_evolution}a).
Reaching very close to the BH, the evolution is dominated by tidal forces,
which lead to a stretching of the build-up cloud in the radial direction and a compression
perpendicular to its motion. Here we define the cloud as the gas atmosphere that accumulates
around S2 due to its wind and the interaction with the surrounding gas.
At this time, Rayleigh-Taylor instabilities and Kelvin-Helmholtz instabilities
have already transformed the shocked-wind region into a filamentary shell (Fig.~\ref{fig:letter_s2star_evolution}b), which is
prone to stripping due to the ram pressure exerted by the accretion flow on the now
fast-moving star ($v_\mathrm{*}^{\mathrm{peri}}\approx 7606\,\mathrm{km}\,\mathrm{s}^{-1}$
at pericentre distance).
A small fraction of these filaments move ahead of the star and have lost enough angular
momentum to the accretion flow to end up inside our inner radius
(Fig.~\ref{fig:letter_densstudy}d), before the nominal pericentre is reached.
During pericentre passage, the typical tidal disruption fan
shows up and the gas that remains bound to the BH smoothes out on a short timescale
caused by the very high sound speed in this region
($c_{\mathrm{s}}^{\mathrm{peri}}\approx 5058\,\mathrm{km}\,\mathrm{s}^{-1}$) and
correspondingly short sound crossing time
($\tau_{\mathrm{sc}}^{\mathrm{peri}}\approx 0.1\,\mathrm{yr}$ compared to
$c_{\mathrm{s}}^{\mathrm{apo}}\approx 1256\,\mathrm{km}\,\mathrm{s}^{-1}$ and $\tau_{\mathrm{sc}}^{\mathrm{apo}}\approx 7.3\,\mathrm{yr}$ at apo centre).
A large fraction of the shell is stripped, but already roughly 0.5\,yr after pericentre
a new shocked wind shell becomes visible. Moving further out in the potential, the radius of this
shell around S2 increases fast (Fig.~\ref{fig:letter_s2star_evolution}c).
Due to the strong hydro instabilities, it forms a thick, filamentary and partly asymmetric
cocoon. Its evolution is similar to the beginning of the simulation, but ram pressure interaction
of the filamentary shocked wind shell with the accretion flow
leads to the formation of a nozzle of gas (Fig.~\ref{fig:letter_s2star_evolution}d) pointing from the
upstream part of the cloud towards the BH and allows to funnel low angular
momentum gas towards the direct surroundings of the minimum of the potential well and
through our inner radius (see discussion in Sect.~\ref{sec:accretion}).
A similar pattern as the one discussed previously starts after having reached
apocentre passage.

\subsection{Mass transfer towards the centre}
\label{sec:accretion}

\begin{figure*}
\centering
\includegraphics[width=\hsize]{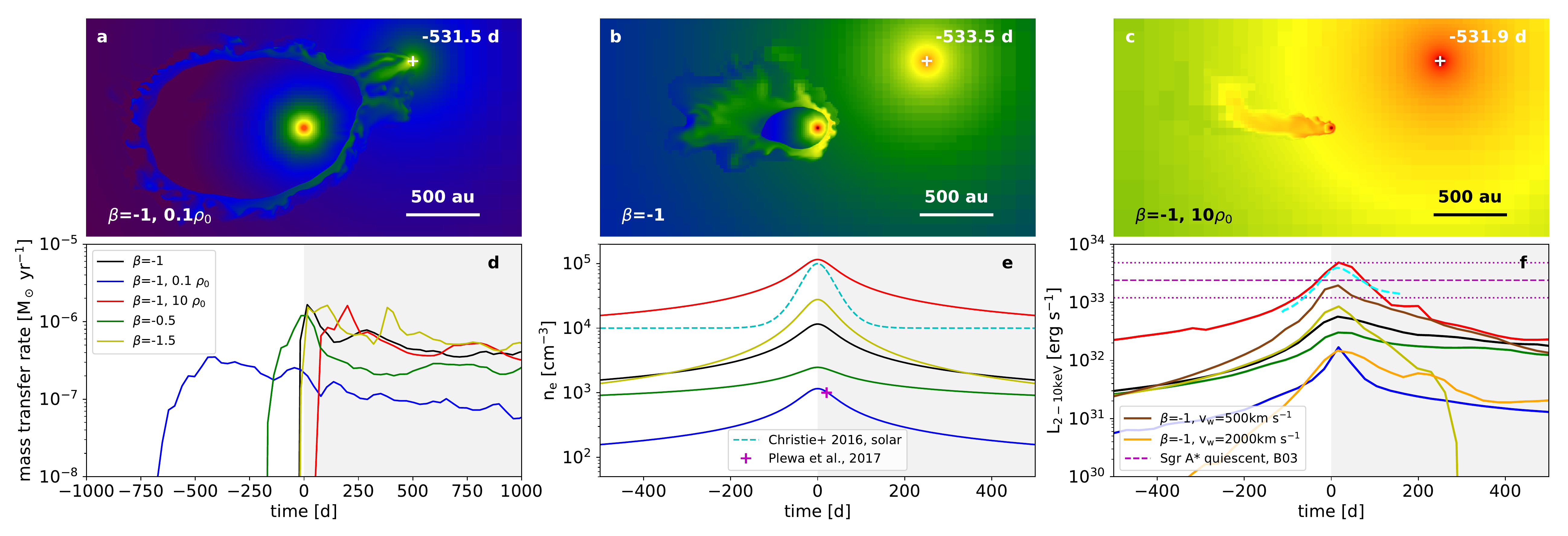}
\caption{Time evolution for a parameter study of various power-law exponents ($\beta$) and
         normalisations ($\rho_0$) of the assumed accretion flow
         density distribution (see legend and Eq.~\ref{equ:densdistribution}), as well as stellar wind velocities $v_{\mathrm{w}}$.
         Shown are cuts through the density in the orbital plane at a time of ~530 days before
         pericentre
         (a-c, same colour scale as in Fig.~\ref{fig:letter_s2star_evolution}a-d; the white cross
         refers to the position of the BH),
         the mass transfer rate through the inner boundary (d),
         the variation of the electron number density of the accretion flow along the orbital path
         of S2 (e), and the light curves of excess X-ray emission (f). Time is given relative to the epoch
         of pericentre passage. The dotted magenta lines denote the accuracy of roughly a factor
         of two of the \citet{Baganoff_03} measurement.
         }
\label{fig:letter_densstudy}
\end{figure*}

The accretion rate (of wind material only, selected by a passive tracer field) through the inner
radius $r_{\mathrm{in}} = 1.4 \times 10^{-4}\,$pc
found in the simulation is shown in Fig.~\ref{fig:letter_densstudy}d
(black line); it peaks close to
pericentre with approximately $2 \times 10^{-6}$\,M$_\sun$ yr$^{-1}$, roughly corresponding to
the upper limit derived from observations at this distance regime \citep{Genzel_10}.
Roughly 1.5\,yr after the violent pericentre passage, the accretion rate reaches an almost
constant level of around $4 \times 10^{-7}$\,M$_\sun$ yr$^{-1}$, as a disc forms around
the BH, which is fed by the stream of gas towards Sgr~A*. The latter is first provided by gas in the dispersed
tail of the source and later by
the nozzle (see Sect.~\ref{sec:dens_evol} and Fig.~\ref{fig:letter_s2star_evolution}) of low-angular-momentum gas connecting BH and source.
Therefore no drastic change in the accretion flow or its electromagnetic emission is expected to occur.

\subsection{X-ray emission}
\label{sec:xray_emission}

Using the recipe described in Sect.~\ref{sec:xray_modelling}, the expected thermal X-ray emission
has been calculated. Being shocked by the interaction with the ambient medium, the stellar wind material
heats up to temperatures of around $10^7$\,K. This gives rise to enhanced thermal X-ray emission
in the 2-10\,keV window compared to our background density structure
(Fig.~\ref{fig:letter_s2star_evolution}e-h), which
roughly follows the morphology of the projected density distribution, but
without showing the free-wind region around the star due to its low temperature of
$T_{\mathrm{fw}}=10^4\,$K.
An increase of the X-ray emission is recognisable the closer the source moves towards the BH. This is caused by (i) the star accelerating into the denser part of the accretion flow
(Fig.~\ref{fig:letter_densstudy}e, black line),
(ii) the continuously increasing amount of gas in the shocked wind shell and
(iii) the tidal interaction. These effects cause stronger shock heating as well as a
stronger compression of the upstream part of the shocked wind shell, leading to higher
densities, temperatures, and emissivity.
In order to derive the corresponding light curve (black line in
Fig.~\ref{fig:letter_densstudy}f), we subtract the X-ray emission
of the accretion flow (our initial condition) and compare to the observed quiescent
emission \citep{Baganoff_03} (dashed magenta line).
We find that the peak of the X-ray emission lags behind the peak of the density evolution
(Fig.~\ref{fig:letter_densstudy}e, black line)
along the orbit. This was already found by \citet{Christie_16} and was explained to be
caused by the compression of the dispersed tail, which reaches the point of maximum
environment density slightly later than the star itself.
The light curve (Fig.~\ref{fig:letter_densstudy}f) is asymmetric with respect to the pericentre passage, showing a
steeper gradient before pericentre compared to the evolution after pericentre.
In the time frame shown, this is caused by the stream
of gas falling back and interacting with the central density concentration and the mixing with high-temperature gas from the accretion flow. For our standard model, the peak (excess) X-ray emission
remains significantly below the observed value (during quiescent phases of Sgr~A*) and no
X-ray flare event is expected.

\subsection{The effect of parameter variations}

In order to determine whether the S2 pericentre passage can be used to constrain the
density distribution of the accretion flow,
we performed an extensive parameter study.
To test the influence of the ambient density on the
stellar wind evolution along the orbit, we decreased and increased the normalisation
of the mean density of the atmosphere
by a factor of ten, in order to bracket the density inferred in \citet{Plewa_17} and
the assumptions made in the simulations of \citet{Christie_16}; see Fig.~\ref{fig:letter_densstudy}e.
Keeping the temperature distribution constant, this changes the
pressure distribution by the same factor and therefore has a strong influence on the
stagnation radius and also on the evolution, as can be seen in Fig.~\ref{fig:letter_densstudy}a-c.
Being able to grow to a larger size in the low-density case (Fig.~\ref{fig:letter_densstudy}a),
the wind-blown bubble reaches the pericentre
earlier and leads to an increased mass transfer rate through our inner boundary
before the nominal pericentre
passage time (Fig.~\ref{fig:letter_densstudy}d).
In contrast, in the high-density ambient medium case (Fig.~\ref{fig:letter_densstudy}c), the
mass transfer increases shortly after pericentre (Fig.~\ref{fig:letter_densstudy}d), but
remains at a similar level to that in our standard model.
In regards to the X-ray emission, the shapes of the light curves behave similarly, whereas the
scaling shows the expected strong differences. The high-density ambient medium case would
seemingly leave an observable X-ray footprint. However, the thermal X-ray emission from the
accretion flow itself would now also exceed the observed quiescent emission, ruling out
this density normalisation.

We change the power-law exponents to -0.5 and -1.5 and normalise the density distributions such that
the thermal X-ray emission within $0.7\,\arcsec$ is given by the quiescent value
of \citet{Baganoff_03}. This results in a factor of 2.4 higher (4.7 lower) density at pericentre
for the case of $\beta=-1.5$ ($\beta=-0.5$, Fig.~\ref{fig:letter_densstudy}e).
As in the previously discussed case, this leads to a shift of the onset of the
mass transfer through our inner boundary
due to the sizes of the shocked wind shell (larger in case of $\beta=-0.5$) and a decrease
(increase) of the X-ray emission for the case of $\beta=-0.5$ ($\beta=-1.5$) due to the
weaker (stronger) compression of the shocked wind shell and the changed ram pressure
interaction with the surrounding medium (Fig.~\ref{fig:letter_densstudy}f).
The sudden drop of the X-ray light curve for the case of $\beta=-1.5$ at around
300 days after pericentre passage is caused by the fast expansion of a dense shell
that was dragged along with the cloud, caused by the steep density (and hence pressure) gradient.
Making the power-law density distribution steeper only
significantly increases the accretion and X-ray emission, if
we fix the density of the accretion flow at the Bondi radius, as has previously been done
in literature. However, this would cause the quiescent thermal emission to deviate from what is observed.
Another source of uncertainty arises from the poorly known parameters of
the stellar wind of S2.
Using the latest measured effective temperature of S2 from \citet{Habibi_17} together with
a fit to stellar wind models of B-type stars (Eq.~1 in \citealp{Krticka_14}), a mass-loss rate of the
order of a few times $10^{-9}$\,M$_\sun$ yr$^{-1}$ is found. The latter is consistent with an
analysis of B-type stars in \citet{Oskinova_11} and questions the values derived in \citet{Martins_08}
(see also discussion in \citealp{Habibi_17}).
A decrease in the wind mass-loss rate is directly correlated with a smaller
stagnation radius and a smaller density in the free
wind region and an overall smaller mass in the cloud.
This allows the cloud to be transformed into a low-density,
thin string of gas resulting in a decrease in the X-ray luminosity
for smaller mass-loss rates.
In a wind velocity study, we find that apart from a slight shift of the initial peak towards later times for
the high-velocity wind ($v_\mathrm{wind}=2000$\,km\,s$^{-1}$), the accretion rates are very similar.
As the stagnation radius scales with the square root of the wind velocity, more compact
wind regions with higher mean densities are expected for lower velocities. This
explains the strong increase of the X-ray emission for our lower wind velocity case ($v_\mathrm{wind}=500$\,km\,s$^{-1}$,
Fig.~\ref{fig:letter_densstudy}f).
Another possibility to increase the X-ray response would be a counter-rotating accretion
flow, which we have not tested so far.

\subsection{Comparison to previous theoretical work}
\label{sec:comp_theo}

The assumed time evolution of the ambient density of the standard model
in \citet{Christie_16} (following an analytical function, supposed to
resemble the crossing of S2 through an accretion disc)
is compared to our study in
Fig.~\ref{fig:letter_densstudy}e. The peak density reached during pericentre passage
is comparable to our model with ambient density increased  by a factor of 10, whereas our
assumed accretion flow structure produces a wider distribution in time compared to the \citet{Christie_16}
time evolution. Astonishingly, these two models result in similar time evolutions
of the X-ray emission, in terms of time offset as well as absolute scaling
(Fig. \ref{fig:letter_densstudy}f). This is surprising because there are many differences
in the setups of the two sets of simulations: different assumptions on the time variation
of the ambient density; a factor of three difference in the assumed mass-loss rate from the star;
two-dimensional vs. three-dimensional simulations; with and
without taking the proper orbit and the tidal interaction into account;
and different numerical resolutions as well as emissivity tables.
By comparing a simulation of the impact of the stellar winds of
roughly 30 Wolf-Rayet stars found in the GC with
and without including the S2 star, \citet{Ressler_18} find that its contribution to the
accretion flow structure is minor, pointing in the same direction as our simulations.

\section{Summary and conclusions}
\label{sec:conclusions}

In this paper, we investigated the interaction of the wind of the S2 star with
the accretion flow in the vicinity of Sgr~A* with the aim of constraining its density
in a so-far unprobed distance regime. We treat the star in isolation and
use an idealised and smooth gas distribution that is in concordance with observations
of the quiescent X-ray emission from Sgr~A*. Compared to previous work, we not only include the
hydrodynamical and thermodynamical, but also the gravitational (tidal) interaction during its orbital
evolution.
By comparing the optically thin, thermal X-ray emission in the 2-10\,keV window
with available observations of the quiescent emission, we find that (given our assumptions) no observable increase of the X-ray emission is expected for the case of our standard model
(even when using a very high upper limit for the stellar wind mass-loss rate).
A significantly higher density of the shocked wind shell close to the pericentre of the orbit
would be required. This could be accomplished by a strong increase of the density
distribution of the simulated accretion flow,
which can, however, be ruled out, as it would exceed the observed quiescent emission.
Only a lower stellar wind velocity leads to an (marginally) observable excess
X-ray emission in our parameter study. This could be boosted by a steeper radial
density profile of the accretion flow, or a broken power law with smaller
exponent in the central region.
Compton-upscattering of optical/UV photons (which we neglect in this letter)
might yield an additional contribution \citep{Nayakshin_05}.
These dependencies on stellar parameters as well as accretion flow parameters lead to a
degeneracy, which does not allow us to constrain the properties of
the accretion flow close to the pericentre distance of the S2 orbit, even if an
X-ray flare were observed during the pericentre passage in May 2018. However, this
might change with the availability of more accurate stellar wind parameters and multi-wavelength
data of this year's pericentre passage.

\begin{acknowledgements}
We thank the referee, Sergei Nayakshin, for useful comments that improved
the quality of the letter as well as D.~Calder{\'o}n, M.~Habibi, and L.~Oskinova for very useful comments and discussion.
For the computations and analysis, we have made use of many open-source software packages,
including {\sc Ramses} \citep{Teyssier_02}, yt \citep{Turk_11}, NumPy, SciPy, matplotlib, hdf5, h4py.
We thank everybody involved in the development of these for their contributions.
MS acknowledges support by the Deutsche Forschungsgemeinschaft through grant no. BU 842/25-1.
The computations were performed on the HPC system HYDRA of the Max Planck Computing and Data Facility.
\end{acknowledgements}

\bibliographystyle{aa} 
\bibliography{literature} 
%

\end{document}